\documentclass[%
reprint,superscriptaddress,amsmath,amssymb,pre,aps,twocolumn,
showkeys
]{revtex4-2}
\usepackage{hyperref}
\usepackage{graphicx}
\usepackage{amsfonts} 
\usepackage{thmbox} 
\usepackage{empheq} 
\usepackage{version}
\usepackage{times}
\usepackage{subfigure}
\usepackage{textcomp}
\usepackage{makeidx}
\usepackage{enumerate}
\usepackage{amsmath}
\usepackage{fancyhdr}
\usepackage[top=20mm, bottom=30mm, left=15mm, right=15mm]{geometry}
\usepackage{color}

\newcommand{\beq}{\begin{equation}}     \newcommand{\eeq}{\end{equation}}
\newcommand{\beqa}{\begin{eqnarray}}    \newcommand{\eeqa}{\end{eqnarray}}
\newcommand{\bde}{\begin{description}}  \newcommand{\ede}{\end{description}}
\newcommand{\ben}{\begin{enumerate}}    \newcommand{\een}{\end{enumerate}}

\newcommand{\la}{\langle}               \newcommand{\ra}{\rangle}

\newcommand{\kT}{{k_{\rm B}T} } 
 
\newcommand{\inv}[1]{{\frac{1}{#1}}}

\newcommand{\inRbracket}[1]{{\left({#1}\right)}}
\newcommand{\inSbracket}[1]{{\left[{#1}\right]}}
\newcommand{\inCbracket}[1]{{\left\{{#1}\right\}}}

\newtheorem[L]{thm}{Theorem}[section]
\newtheorem{cor}[thm]{Corollary}
\newtheorem{theorem}{{\sf Assertion :}}[section] 
\newtheorem{definition}{\sf Definition} 
 
\newtheorem{lemma}[theorem]{Lemma}
\newcommand{\bth}{\begin{thm}}  
\newcommand{\blem}{\begin{lemma}}  
	\newcommand{\elem}{\end{lemma}}  
\newcommand{\bpr}{\begin{proof}}  
	\newcommand{\epr}{\end{proof}} 
\newcommand{\bdefine}{\begin{definition}}  
	\newcommand{\edefine}{\end{definition}}  
\newcommand{\bcor}{\begin{cor}} 
	\newcommand{\ecor}{\end{cor}}  
\newcommand{\bprop}{\begin{example}[Property]}  
	\newcommand{\eprop}{\end{example}}  
\newcounter{formulaire}
\newcommand{\beqf}{\addtocounter{formulaire}{1}\begin{equation}}
	\newcommand{\eeqf}{\tag{R \arabic{formulaire}}\end{equation}}
\newcommand{\beqaf}{\addtocounter{formulaire}{1}\begin{equation}\begin{array}{rcl}}
		\newcommand{\eeqaf}{\end{array}\tag{R \arabic{formulaire}}\end{equation}}

\newcommand{\veca}{{\vec{a} } }
\newcommand{\vecx}{{\vec{x}}}
\newcommand{\vecabs}[1]{{\| \vec{{#1}}\|}}

\usepackage{dcolumn}
\usepackage{bm}
\usepackage{listings}
\usepackage{ulem}
\usepackage{graphicx}
\usepackage{subfigure}

\begin{document}
\title{Martingale drift of Langevin dynamics and classical canonical spin statistics  \texorpdfstring{\\}{} }
\author{Ken Sekimoto}
\affiliation{Laboratoire Mati\`ere et Syst\`emes Complexes, UMR CNRS 7057, Universit\'e Paris Cit\'e,\\
		10 Rue Alice Domon et Léonie Duquet, 75013, Paris, France }
\affiliation{Laboratoire Gulliver, UMR CNRS 7083, ESPCI Paris, Universit\'e PSL\\
		10 rue Vauquelin, 75005, Paris, France.}
\email[Corresponding author: ]{ken.sekimoto@espci.fr}
	
\date{ \today
} 
	 
\begin{abstract} 
A martingale is a stochastic process that encodes a kind of fairness or unbias, which is associated with a reference process. Here we show that, if the reference process $x_t$ evolves according to the Langevin equation with drift $a(x),$ and if $a(x_t)$ is martingale, then its amplitude is the Langevin function, which originally described the canonical response of a single classical Heisenberg spin under static field. Furthermore, the asymptotic limit of $x_t/t$ obeys the ensemble statistics of such Heisenberg spin.
\end{abstract}
\maketitle
\section{\null{Backgrounds}}\label{sec:intro}
Paul Langevin derived in 1905 what we call nowadays the Langevin function \cite{langevinfunction} that gives the canonical equilibrium response of paramagnetism under static magnetic field. It was based on the Boltzmann-Einstein statistics and explained the Curie's law, which Pierre Curie has established experimentally ten years before \cite{curieLaw}. 
In the modern language, a classical 3D Heisenberg spin  under a uniform magnetic field along an axis undergoes the thermal fluctuation according to the Boltzmann weight, or the relative probability, $e^{x\cos\theta},$ where $\theta$ is the polar angle of the spin relative to the field axis and $x$ is the strength of the magnetic field scaled by the temperature and magnetic moment of the spin. The mean polarisation of the spin is given by the average of $\cos\theta$ weighted  by the above probability and integrated over the solid angles. The result is the Langevin function, $\coth{x}-\inv{x}$ \cite{langevinfunction}. That it behaves like $\inv{3}x$ for small $|x|$ explained the Curie's law.

Three years after Albert Einstein \cite{Einstein1905} had formulated Guoy's qualitative idea of Brownian motion \cite{Gouy1,Gouy2}, P. Langevin introduced in 1908 the first stochastic differential equation that is nowadays called the Langevin equation \cite{langevinpaper}. 
In the modern language a simple and generic form of Langevin equation in $d$-dimension reads
\beq\label{eq:eqL}
\frac{d{\vec{x}_t}}{dt}=\veca({\vec{x}_t})+\vec{\xi}_t,
\eeq 
which contains the {drift}, $\veca(\vecx_t),$ and the white Gaussian noise, $\vec{\xi},$ with the zero mean and the unit diagonal covariance, $\langle \vec{\xi}_t,\vec{\xi}_s\rangle
=\bm{1}\delta(t-s),$ where $\bm{1}$ is the unit tensor. 
While we nowadays know how this equation admits the first \cite{sekimoto97,sekimoto98} and second \cite{udo,udo-review2012} laws of thermodynamics, it was Langevin who revolutionarized the notion of the evolution equation as a mapping between the path ensembles.
In the 1900's, however, no link between the Langevin function and Langevin equation has been known, to the author's knowledge.

It is around 1940 that Jean Viller and Doob modernized the historically old notion of martingale as a powerful concept of the modern probability theory, see for example the Chap.1 of \cite{review300} for a historical review.
The martingale is the mathematical expression of the idea of the fairness for the future:
The continuous-time stochastic process $\vec{Y}_t$ is said to be martingale associated with the stochastic process $\{\vec{X}_{s}\}$ if {\it (i)} $\vec{Y}_t$ is causally determined by $\{\vec{X}_{s}\}_{0\le s\le t}$ and {\it (ii)} its conditional expectation for future shows the fairness:
\beq\label{eq:mtgl}
\langle \vec{Y}_t| \{\vec{X}_{u}\}_{0\le u\le s}\rangle=\vec{Y}_s\quad \mbox{for $\forall t\ge s$},
\eeq
let alone the other rigorous mathematical conditions. Here 
$ \{\vec{X}_{u}\}_{0\le u\le s}$ denotes the history of $\vec{X}_u$ over the peiode $0\le u\le s$ and $\la R |{\it cond}\ra$ means the expectation of the random variable $R$ under the condition, {\it cond}.
For the later convenience we introduce the differential version
which follows from (\ref{eq:mtgl}) applied to $Y_{t+dt}$ as well as to $Y_t$:
\beq\label{eq:mtgl-diff}
\langle d\vec{Y}_t| \{\vec{X}_{u}\}_{0\le u\le s}\rangle=0 \mbox{ for } \forall t\ge s,
\eeq
As the probability theory provides with many useful theorems about martingale, efforts have often been made to convert a reference process of interest into a martingale process  to draw new properties of $X_t.$ The population dynamics or mathematical finances have developed that approach since longtime \cite{review300}.
In the stochastic thermodynamics people recently recognized that the exponentiated entropy production is the martingale process of the category called path-probability ratio or Radon-Nikodym density process \cite{martingale-Gupta2011,Roldan-prX2017}. 

A distinct type of martingale other than the path-probability ratio has also been found in what we call Progressive Quenching \cite{PQ-KS-BV-pre2018,PQ-chain,PQ-CM-KS-2020,PQ-CM-KS-2022}. Starting from an equilibrium many spin system, we fix, one after another, the spins at the instantaneous state they take. Through such an unbiased protocol, the martingale process was found not in the progressively fixed spins, but in the mean spin {\it to be fixed.} Those already fixed spins influence on the latter through the molecular field. We have coined a name ``hidden martingale'' for it because, in the continuous-time counterpart, it is not $\vecx_t$ in (\ref{eq:eqL}) but the {drift} $\veca(\vecx_t)$ that should be martingale --- This is the very starting point of the present study. 
Below, in Sec.\ref{sec:Lfunc} we show that, if the {drift} $\veca(\vecx_t)$ in (\ref{eq:eqL}) is martingale associated with the process (\ref{eq:eqL}) itself, then the {drift} is the Langevin function\footnote{The contents essentially up to Sec.\ref{sec:Lfunc} is accessible as arXiv:2305.04976v1.}. The emergence of a spin-related function is surprising because the Langevin equation (\ref{eq:eqL}) works in a non-compact space. However, a further surprise in Sec.\ref{eq:miroscope} is that the canonical statistics of such spin emerges in the ensemble of the long-time asymptotes of $x_t/t.$

\section{Langevin function as self-harmonic drift}\label{sec:Lfunc}
\subsection*{IIA.\,Family of harmonic functions associated with a drift} 
When the process $\vec{Y}_t=\vec{h}(\vec{X}_t)$ is martingale associated with the process $\{\vec{X}_{s}\}$, the function $\vec{h}$ is said to be harmonic \cite{review300}. In particular, if the process $\{\vec{X}_{s}\}=\{\vecx_t\}$ is generated by (\ref{eq:eqL}) and that $\vec{h}(\vecx_t)$ takes the value in the same space as $\vecx_t$ does, the condition for the harmonicity reads
\beq \label{eq:harmh}
(\veca\cdot\nabla)\vec{h}+\inv{2}\Delta\vec{h}=0.
\eeq
To see this it suffices to develop $d\vec{h}(\vecx_t)$ up to the order $\mathcal{O}(dt)$, that is, 
\beq\label{eq:dh2dt}
d\vec{h}(\vecx_t)=(d\vecx_t\bullet \nabla)\vec{h}+ \inv{2}(d\vecx_t\bullet \nabla)^2\vec{h},
\eeq 
where $\bullet$ is the It\^o-type product \cite{review300}.
We substitute into it the stochastic differential version of (\ref{eq:eqL}), 
\beq\label{eq:eqSDE}
d\vecx_t=\veca(\vecx_t)dt+d\vec{W}_t,
\eeq
where $\vec{W}_t$ is the $d$-dimensional standardized Brownian motion, i.e., 
the Wiener process, such that $d\vec{W}_t d\vec{W}_t = \bm{1}dt$ with $\bm{1}$ being the unit tensor.
The imposition of the martingale condition (\ref{eq:mtgl-diff}) on $\vec{Y}_t=\vec{h}(\vecx_t)$ then leads to (\ref{eq:harmh}).
The form  (\ref{eq:harmh}) explains the denomination `harmonic' because, 
in the absence of the {drift}, $\veca\equiv 0,$ the Eq.(\ref{eq:harmh}) is the vector Laplace equation.

Given a drift $\veca$ we can conceive the family of harmonic functions associated with the process obeying (\ref{eq:eqL}), We shall denote this family by $\mathcal{H}_{\veca}:$ 
\beq
\mathcal{H}_{\veca}=\inCbracket{\vec{h}\,\, ; \,
(\veca\cdot\nabla)\vec{h}+\inv{2}\Delta\vec{h}=0}.
\eeq

\subsection*{IIB.\,Langevin function as fixed point} 
Among the families associated with different drift, the self-referential condition
\beq	\label{eq:fixedpt}
{\veca^{\,*}}\in \mathcal{H}_{\veca^{\,*}}
\eeq
defines the special drift $\veca^{\,*}$ as a kind of fixed point, which we shall call the {\it self-harmonic} drift when it exists. More concretely, 
\beq\label{eq:key}
({\veca^{\,*}}\!\cdot\!\nabla){\veca^{\,*}}+\inv{2}\Delta {\veca^{\,*}} =\vec{0}.
\eeq
This is the key equation of the present paper.

There is possibilities of asymmetric solutions of (\ref{eq:key}). Apparently the cylindrical lift-up of solution for the lower-dimensions is an anisotropic solution. Nevertheless, we here seek for the solutions which are isotropic with respect to the origin, $\vecx=0.$ ({\it Note:} The process $\vecx_t$ can nevertheless start from any $\vecx$ off the origin.)

We then assume the form\footnote{We use the notation $\vecx$ rather than $\vec{r}$ because it is generally not be a spatial position and, moreover, $L(x)$ is more common than $L(r)$ in the literature.}:
\beq\label{eq:isotropic}
\veca^{\,*}(\vecx)= L_d(\vecabs{x})\hat{x},
\eeq
where $\hat{x} \equiv {\vecx}/{\vecabs{x}}$ is the unit vector along $\vecx$ and the suffix $d$ stands for the space dimension. 
Then (\ref{eq:key}) implies 
\beq\label{eq:LdODE}
L_d''(x)+2 L_d(x) L_d'(x)+\frac{d-1}{x}\inRbracket{L_d'(x)-\frac{L_d(x)}{x}}=0.
\eeq
In the two-dimensional family of solutions of (\ref{eq:LdODE}) the invariance under the similarity transformation, $L_d(x)\to \alpha L_d(\alpha x)$ provides with one parameter. If we write the first integral of (\ref{eq:LdODE}) as 
\beq\label{eq:Riccati} 
L_d'(x)+(L_d(x))^2+\frac{d-1}{x} L_d(x)=\alpha^2, 
\eeq
then we can reduce the problem to the finding of $L_d(x)$ for (\ref{eq:Riccati}) with $\alpha=1.$ 
Then in the remaining one-parameter family of solutions for (\ref{eq:Riccati}),
some analysis using Mathematica$^{\mbox{\textregistered}}$ tells that, at least for $d=$2,3 and 4, there is a unique solution which does not diverge at $x=0.$
For example in $d=3,$ the solution $L_3^{(\beta)}(x)\equiv  \coth(x+\beta)-\inv{x}$ is regular only with $\beta=0.$
For $d=1$ we can solve (\ref{eq:key}) directly so that $\alpha=1$ in (\ref{eq:Riccati}) and choose appropriately the origin.
Altogether, the solutions thus identified are as below:
\beqa\label{eq:Ld}
L_1(x)&:=&\tanh(x)
\cr 
L_2(x)&:=& \frac{I_1(x)}{I_0(x)}
\cr
L_3(x)&:=&\coth{x}-\inv{x}
\cr 
L_4(x)&:=&\frac{x [I_0(x)+ I_2(x)]-2 I_1(x)}{2 x I_1(x)},
\eeqa
for $x\neq 0$ and $L_d(0)=0$ for any dimension, $d,$
where $I_n(x)$ are the $n$-th modified Bessel functions of the first kind.
A standard but inspiring approach to obtain these solution from (\ref{eq:Riccati}) with $\alpha=1$ is to 
use the (Riccati) transformation,
\beq\label{eq:RiccatiTr}
L_d\equiv \frac{Z'_d}{Z_d} =\frac{d}{dx}[\ln Z_d(x)],
\eeq
which renders (\ref{eq:Riccati}) into the linear equation,
\beq\label{eq:ODEforZd}
\frac{d^2 Z_d}{dx^2}+\frac{d-1}{x}\frac{d Z_d}{dx}= Z_d.
\eeq
By noticing the radial part of the Laplacian operator on the l.h.s. of (\ref{eq:ODEforZd}),
we find \beq \label{eq:ZasPF}
Z_d(\| \vecx\|)\propto \oint_{\|\hat{S}\|=1}e^{\hat{S}\cdot\vecx}d\Omega_S,
\eeq
that is, the partition function for a single classical Heisenberg spin under the non-dimensionalized external field, $\vecx.$ 
Through (\ref{eq:RiccatiTr}) the drift is, therefore, the canonical average of the spin $\hat{S}$ under this field.
All $L_d(x)$'s are odd in $x$ and their graphs look similar to the simplest one, $L_1(x),$ or the original Langevin function, $L_3(x).$  They also have the unique limit, $\lim_{x\to\infty}L_d(x)=1.$ We may call $L_d(x)$ the {\it $d$-dimensional Langevin function.} In any case the Langevin equation meets here with the Langevin functions in the context of the martingale.
For completeness we list below $Z_d(x)$ with the numerical coefficient so that $Z_d(0)$ is the surface area of $d$-dimensional unit hyper-sphere: 
\beqa \label{eq:Zs}
&&(Z_1,Z_2,Z_3,Z_4)
\cr &&=\inRbracket{2\cosh(x),2\pi I_0(x), 4\pi\frac{\sinh(x)}{x},4\pi^2\frac{I_1(x)}{x}}.
\eeqa

\paragraph{Remark on the solution family} --- 
The change of the units of length and time causes the rewriting of (\ref{eq:eqSDE}).
If we introduce $\vec{y}$ and $\tau$ through $\vecx=\alpha \vec{y}$ and $t= \alpha^2 C\tau,$ then (\ref{eq:eqL}) with (\ref{eq:isotropic}) for $\veca(\vecx_t)$ leads to
\beq\label{eq:scale}
 d\vec{y}_\tau = C \alpha\, L_d(\alpha{\|\vec{y}_\tau\|}{}) \hat{y}_\tau  d\tau + \sqrt{C} d\vec{W}_\tau,
\eeq
where $\hat{y}_\tau\equiv \vec{y}_\tau /\|\vec{y}_\tau\| $ and
we have used the statistical equivalence, $d\vec{W}_{\kappa t}\simeq \sqrt{\kappa} \,d\vec{W}_t,$ for the standard $d$-dimensional Wiener process. In (\ref{eq:scale}) we see that, apart from the flexibility of the self-harmonic function, $\alpha L_d(\alpha{\|\vec{y}_\tau\|}{}),$ 
 there is a specific scale relationship between the drift and diffusion for the drift to be martingale.
See Appendix \ref{app:scaling} for more argument. 

\section{Statistical property of the asymptotic limits : ``microscope'' }\label{eq:miroscope}
In general the Langevin functions $L_d(x)$ can appear in being unrelated with the canonical statistics of a Heisenberg spin.
For example, $L_2(x)$ has appeared as the 1+1-dimensional nucleation-controlled polymer crystal growth rate \cite{Bennett-Butt-Land1981,Doi-Peliti-Goldenfeld1984}, with no relevance to the XY-spin.
In this section we show that the statistics of a Heisenberg spin indeed appears in the asymptotic behavior of the Langevin dynamics with self-harmonic drift.

\subsection*{IIIA.\,Convergence of the self-harmonic drift and asymptotic behavior of Langevin process}
Since the velocity $d\vecx_t/dt$ is on the average oriented along $\vecx_t$ (see (\ref{eq:isotropic})) and that $L_d(\|\vecx_t\|)$ is nonnegative for any nonzero $\vecx_t,$ we may expect that $\vecx_t$ most probably grows unboundedly for a long time. However, its further detail is not clear at first glance. We, therefore, focus first on the evolution of the {drift}, $\veca^{\,*}(\vecx_t).$ 
Because of the martingale condition (\ref{eq:key}) the development of 
$d\veca^{\,*}(\vecx_t)$  contains no (drift) term for $\mathcal{O}(dt)$ (see (\ref{eq:dh2dt})),  
and the result reads 
\beq\label{eq:aevol}
d\veca^{\,*}(\vecx_t)=  d\vec{W}_t\bullet (\nabla \veca^{\,*}) ,
\eeq
\beq\label{eq:nablaveca}
\nabla \veca^{\,*}= \frac{dL_d(\chi)}{d\chi} \hat{a}^*\hat{a}^* 
+  \frac{L_d(\chi)}{\chi} (\bm{1}-\hat{a}^*\hat{a}^* ),
\eeq
 where $\hat{a}^*\equiv  \veca^{\,*}/\|\veca^{\,*}\|$ and $\chi$ is 
 inversely determined so that $L_d(\chi)=\|\veca^{\,*}\|.$ Thus $(\nabla \veca^{\,*})$ is a function of $\veca^{\,*}.$ The process generated by $d\veca_t =d\vec{W}_t\bullet {\sf M}(\veca_t)$ with ${\sf M}(\veca)$ being any rank-2 tensor as function of $\veca$ is martingale associated with the Wiener process, $\vec{W}_s$  $(0\leq s<\infty).$ The self-harmonic drift $\veca^{\,*}$ defined above, however, has several particular properties on top of it, which we will discuss below.

We notice that $\veca^{\,*}(\vecx_t)$ of each realisation converges for $t\to\infty$ and the limit $\veca^{\,*}_\infty\equiv \lim_{t\to\infty}\veca^{\,*}(\vecx_t)$ is on the hypersphere, $\|\veca^{\,*}_\infty\|=1.$
The case of $d=1$ is particularly simple: Eqs.(\ref{eq:aevol}) and (\ref{eq:nablaveca}) reads
\beq\label{eq:1Da}
da^*_t=(1-{a^*_t}^2)\bullet dW_t
\eeq
because $(\tanh x)'=1-\tanh^2 x.$
The Fokker-Planck (FP) equation equivalent to (\ref{eq:1Da}) has the potential, $\ln[(1-{a^*}^2)^2]$ and the diffusion coefficient  $\inv{2}(1-{a^*}^2)^2.$ The variable $\veca^{\,*}_t$ should be pushed towards $\pm 1$ and stuck there.
In fact the singular densities $\delta(a_\infty^*\mp 1)$ or their linear combinations are all stationary solutions of that FP equation. 

While the behavior of $\veca^{\,*}(\vecx_t)$ for $d>1$ is less clear than for $d=1,$ 
the fact that $\nabla a^*$ vanishes on the hyper-sphere $\|\veca^{\,*}\|=1$ (i.e. at $\chi=+\infty$) indicates that the hyper-sphere is the natural absorbing boundary for the evolution of $\veca^{\,*}(\vecx_t).$
More formal and general argument for the existence of limit is provided by the {\it convergence theorem of (sub)martingale} (see, for e.g., \S4.1.5 of \cite{review300}). According to this theorem, because $\veca^{\,*}(\vecx)$ is bounded, i.e., $\|\veca^{\,*} \|\le 1,$ the martingale process 
$\veca^{\,*}(\vecx_t)$ should have a limit; 
\beq\label{eq:ainfBYa}
\veca^{\,*}_\infty\equiv \lim_{t\to\infty}\veca^{\,*}(\vecx_t).
\eeq
On the other hand, if the limit $\veca^{\,*}_\infty$ were not on the natural boundary, the noise for $\vecx_t$ could still relocate it. Therefore, $\|\veca^{\,*}_\infty\|=1$ is concluded.

\begin{figure}[h]
\includegraphics[width = 0.9 \linewidth]{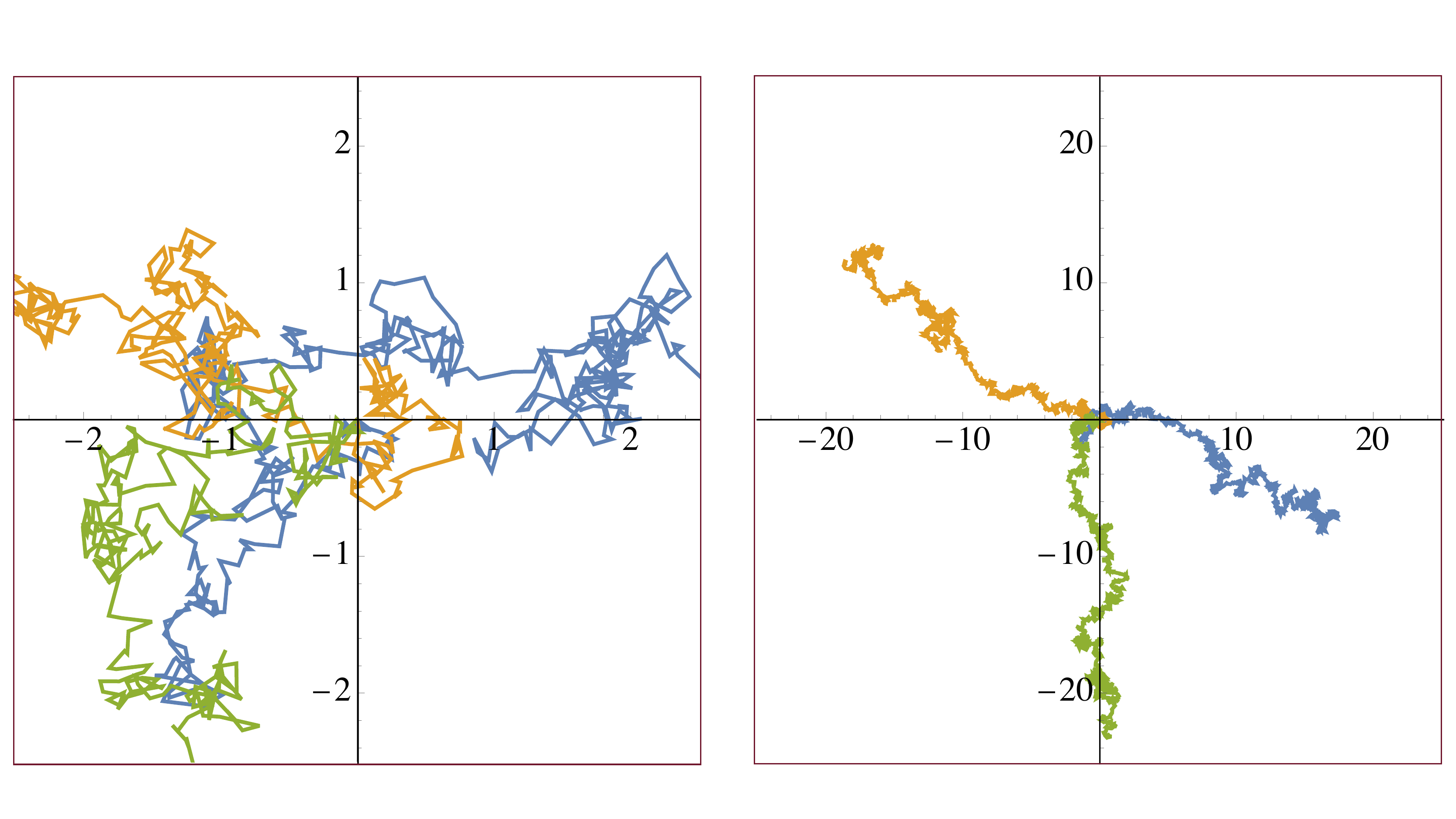}
\caption{
{2D trajectories from $\vecx_0=(0_+,0_+).$ (Left) Close-up view near the origin. (Right) Whole trajectories up to $t=16.$ }
}\label{fig:2scales}
\end{figure}

Next we consider the trajectory of $\vecx_t$ that leads to $\lim_{t\to \infty}\veca^{\,*}(\vecx_t)=\veca^{\,*}_\infty$ ? 
Fig.\ref{fig:2scales} shows three samples in $d=2,$ having started from the origin, $\vecx_0=0.$
The left figure is the close-up view of the initial part of the trajectories and the right figure shows the entire trajectory up to $t=16$ including the left figure\footnote{The $0_+$ is a numerical artificial infinitesimal, being used to avoid the removable singularity of $L_d(x)$ at $x=0$}.
In the early stage with  $\|\vecx_t\|\lesssim 1,$ the noise term $\vec{\xi}_t$ dominates over the {drift}, then there is a slow cross-over to the long-time ballistic behavior.
This observation together with (\ref{eq:eqL}) or (\ref{eq:eqSDE}), we understand that 
$\vecx_t$ becomes asymptotically ballistic in the sense that 
\beq\label{eq:ainfBYx}
\lim_{t\to\infty} \frac{\vecx_t}{t}= \veca^{\,*}_\infty,
\eeq
which also imply $\lim_{t\to\infty} \frac{\|\vecx_t\|}{t}= 1.$
In Appendix \ref{app:convergence} we show (\ref{eq:ainfBYx}) in more detail.
Note that each realisation of trajectory ends up with a particular orientation
of $\veca^{\,*}_\infty.$

\subsection*{IIIB.\, Distribution of trajectories at $t\to\infty$ as ``spin'' microscope}
\begin{figure}[h]
\centering
\subfigure[$x_0=0$]
{\label{subfig:X0eq0}    \includegraphics[width=4.2cm,angle=0.]{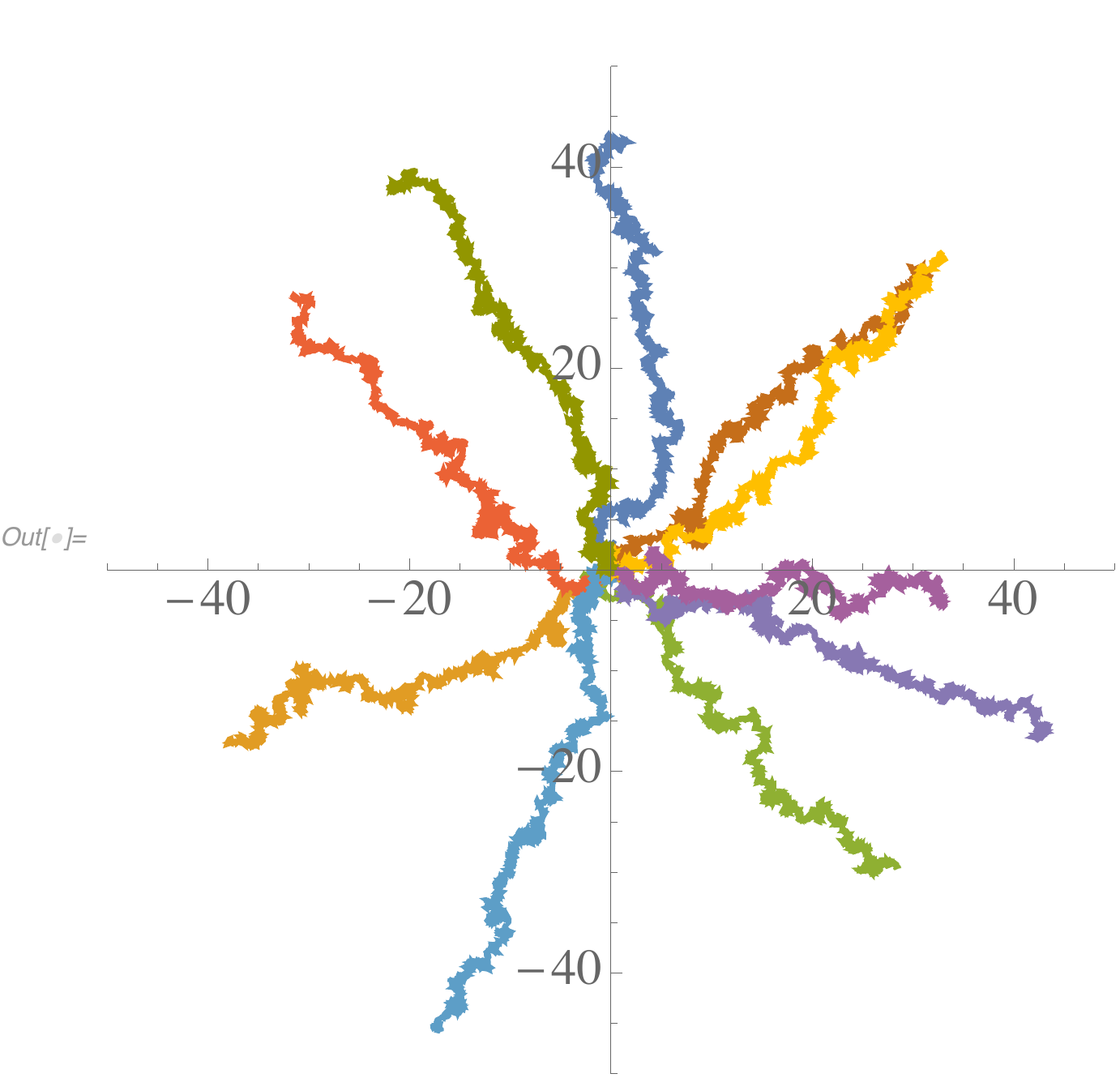}} 
\hspace{2mm}
\subfigure[$x_0=1$]
{\label{subfig:X0eq1}    \includegraphics[width=4.2cm,angle=0.]{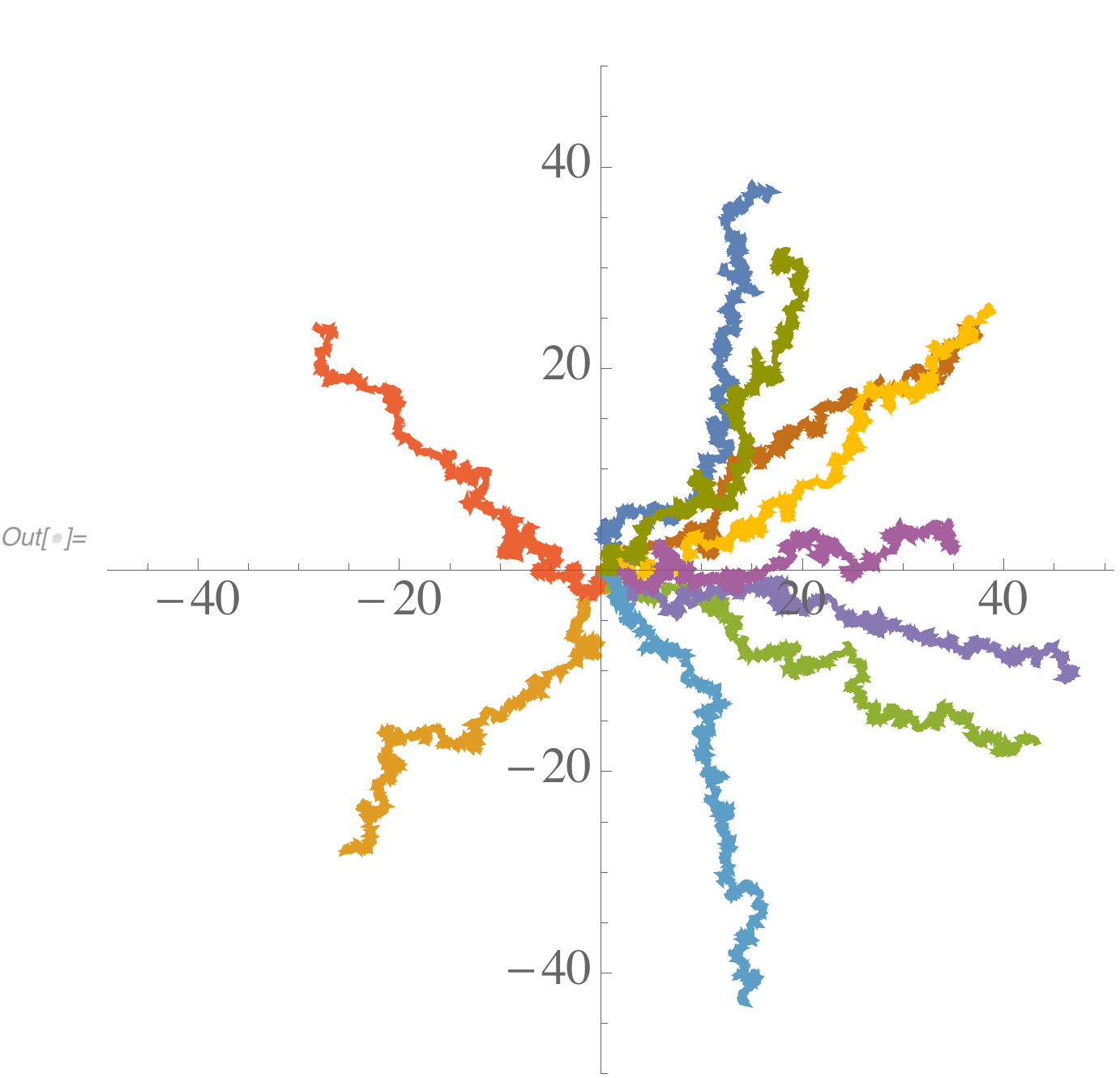}} 
\hspace{2mm}
\subfigure[$x_0=2$]
{\label{subfig:X0eq2}    \includegraphics[width=4.2cm,angle=0.]{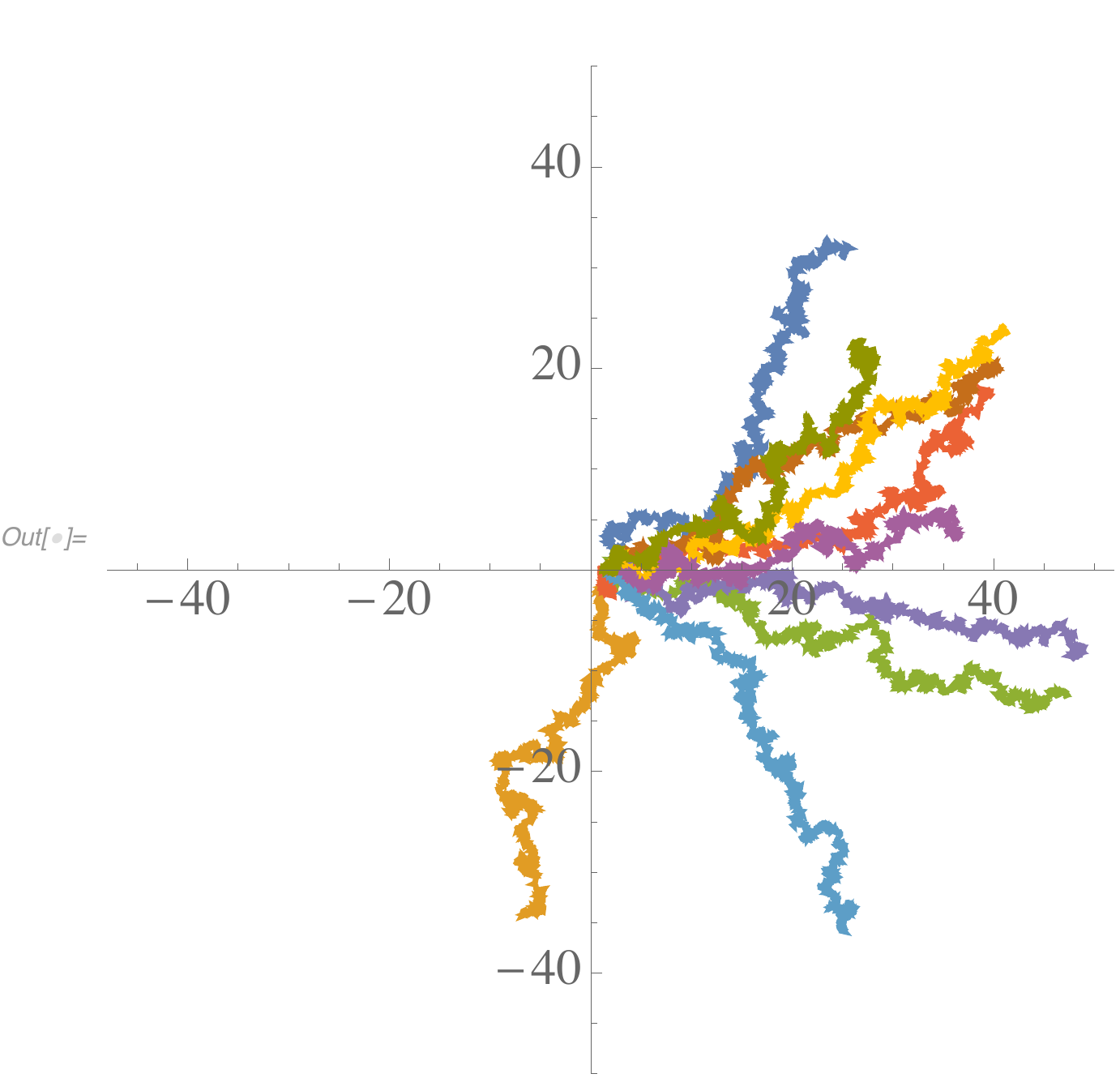}} 
\hspace{2mm}
\subfigure[$x_0=4$]
{\label{subfig:X0eq4}    \includegraphics[width=4.2cm,angle=0.]{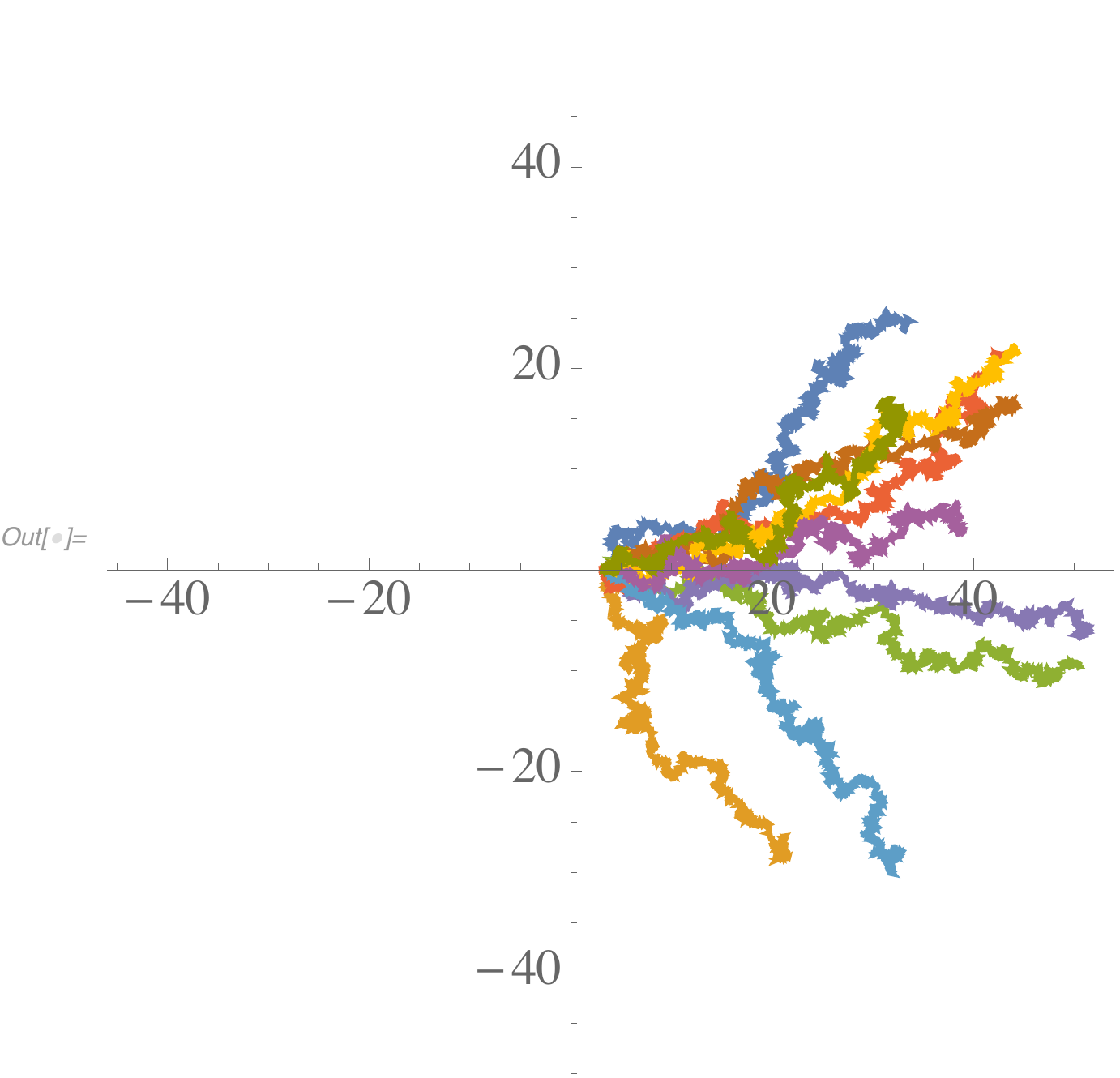}} 
\caption{2D trajectories from $\vecx_0=(x_0,0_+)$ with (a) $x_0=0$, (b) $x_0=1$, (c) $x_0=2$ and (d) $x_0=4.$ In all cases the duration is up to $t=40$ and 
we used the identical set of noise histories for $\vec{\xi}_t$ in (\ref{eq:eqL})
}
\label{fig:traj} 
\end{figure}

Each trajectory of $\vecx_t$ or of $\veca^{\,*}(\vecx_t)$ has a the limiting value $\veca^{\,*}_\infty$ as random variable and, by the definition of self-harmonicity, we have the martingality:
\beq \label{eq:mtglDainf}
\la \veca^{\,*}_\infty|\vecx_0\ra= \veca^{\,*}(\vecx_0).
\eeq
This is a constraint on the statistics of $\veca^{\,*}_\infty.$
However, it is only for $d=1$ that the probability of realizing $a_\infty^*=\pm 1$ and 
the initial data $a^{*}(x_0)$ are trivially related by (\ref{eq:mtglDainf}),
$$
Prob(a_\infty^*=s)=\frac{1+s\,a^{*}(x_0)}{2},\quad s=\pm 1.
$$
By contrast, the statistics of $\veca^{\,*}_\infty$ on the $d$($>1$)-dimensional hyper-sphere surface is a by no means trivial.
Fig.\ref{fig:traj} gives an qualitative idea of how the trajectories 
of $\vecx_t$ up to $t=40$ depends the initial data, $\vecx_0.$ 
Roughly, the larger is $\|x_0\|,$ the more polarized is the orientation of $\vecx_t.$ 

We shall parameterize the orientation of $\veca^{\,*}_\infty$ 
by its orthogonal projection, $\cos \theta\equiv \veca^{\,*}_\infty\cdot(\vecx_0/ \|\vecx_0\|),$ onto the axis along $\vecx_0,$
where $0\le \theta\le \pi.$
When $\vecx_0=0,$ the distribution of $\veca^{\,*}_\infty$ through  (\ref{eq:ainfBYx}) must be isotropic and the cumulative probability distribution of $\cos\theta$ should rigorously obey 
$Prob^{\rm (can)}(\cos\theta<\chi)=\inv{\pi}\int_{-1}^\chi \frac{d\xi}{\sqrt{1-\xi^2}}
=1-\inv{\pi}\arccos\chi.$
\begin{figure}[t!!] %
\includegraphics[width = 0.8 \linewidth]{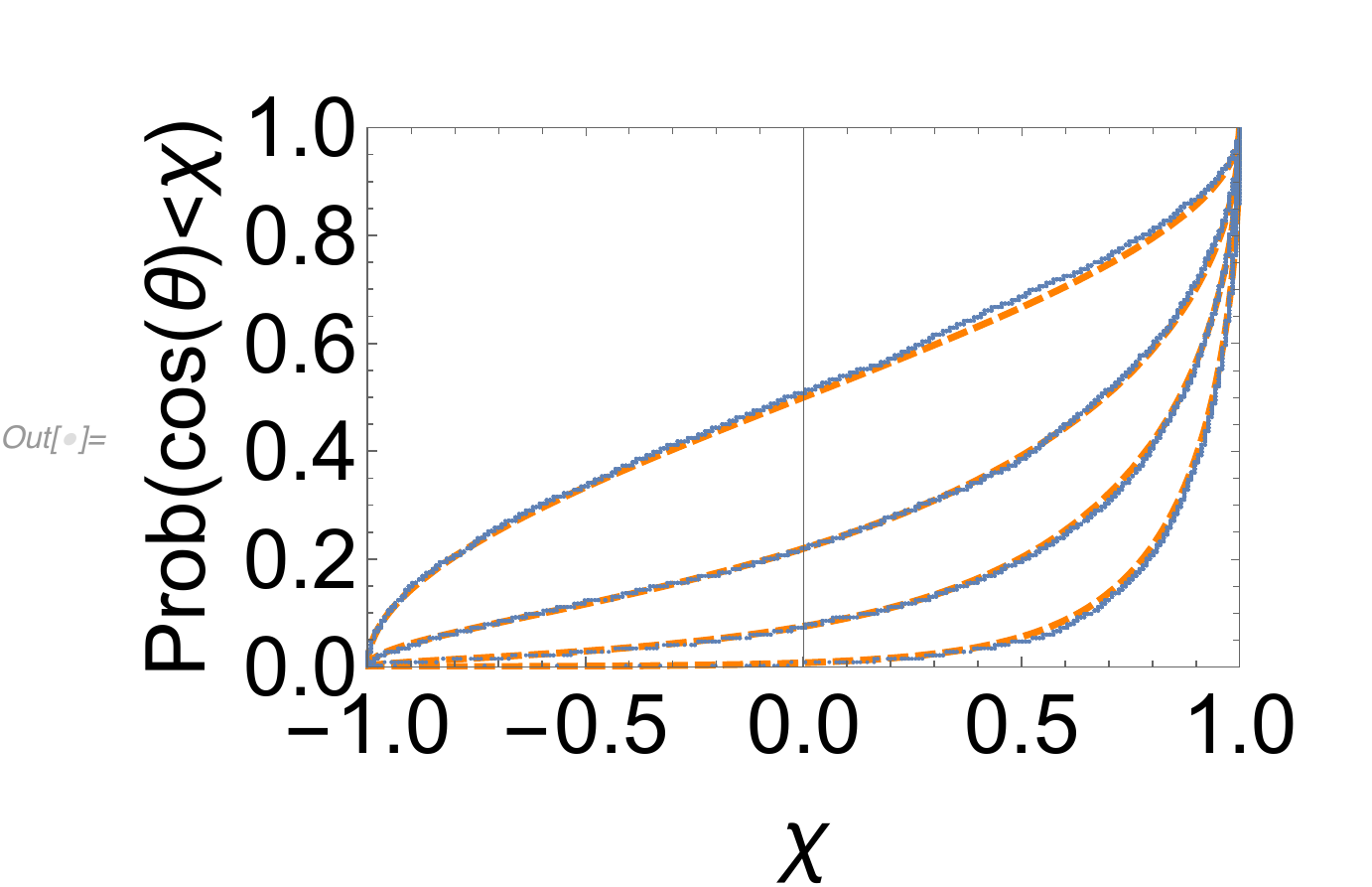}
\caption{{The 2D orientational distributions of $\lim_{t\to\infty}(\null{\vecx_t}/{t})=\veca^{\,*}_\infty$ for the different starting points, $x_0=0,1,2,$ and $4,$ from top to bottom. 
 We represent the orientation of $\veca^{\,*}_\infty$ using $\cos\theta,$ where $\theta$ is the angle between $\veca^{\,*}_\infty$ and $\vecx_0=(x_0,0),$ 
 and we show the distributions by the empirical cumulative probability. 
 Numerically the ensemble of $\veca^{\,*}_\infty$ is approximated by 3000 realisations of $\vecx_t/t|_{t=40}.$ 
 The red dashed curves represent the canonical equilibrium distribution of a single unitary spin under the non-dimensionalized field $\vecx_0,$ 
 see (\ref{eq:cancumul}) in the main text. }
}\label{fig:magic}
\end{figure}
In Fig.\ref{fig:magic} the top dashed curve (in red) represents this formula.
The train of dots (in blue) along this curve are the results obtained from the numerical data over 3000 trajectories, where $\veca^{\,*}_\infty=\lim_{t\to\infty}\frac{\vecx_t}{t}$ is approximated by $\left.\frac{\vecx_t}{t}\right|_{t=40}.$
 The deviations from the theoretical curve shows the errors due to the finiteness of sampling. The other three trains of blue dots in Fig.\ref{fig:magic} represent the numerically obtained cumulative probabilities $\cos\theta$ for $x_0=1,2$ and $4,$ respectively, in the descending order. 
({\it Note:} We took the cumulative probability because it can be empirically reconstructed just by plotting the normalized rank $r/3000$ vs the corresponding value of $\cos\theta_r,$ without any binning or smoothing.)

Quite surprisingly, the results almost surely obey the canonical statistics of a Heisenberg spin under the non-dimensonalized field, $\vecx_0=(x_0,0),$ which leads to the cumulated probability,
\beq \label{eq:cancumul}
Prob^{\rm (can)}(\cos\theta<\chi |\vecx_0)=\inv{\pi I_0(x_0)}\int_{-1}^\chi \frac{e^{x_0 \xi}d\xi}{\sqrt{1-\xi^2}}.
\eeq
This formula is shown by the dashed curves (in red) in Fig.\ref{fig:magic}.
Thus the thermal distribution of the (fictitious) spin orientations that would yield the expectation $\veca^{\,*}(\vecx_0)$ has been mapped into the static distribution of the ``spin,'' $\veca^{\,*}_\infty,$ characterizing the trajectory of $\vecx_t$ in the limit of $t\to\infty.$ Being consistent with (\ref{eq:mtglDainf}), our proposition for the probability density $\rho\inRbracket{\veca^{\,*}_\infty}$  is, therefore, 
\beq \label{eq:propo}
\rho\inRbracket{\veca^{\,*}_\infty}=
\frac{e^{\vecx_0\cdot \veca^{\,*}_\infty}}{Z_d(\|\vecx_0\|)},
\qquad \veca^{\,*}_\infty\in S^{d-1},
\eeq
where $\veca^{\,*}_\infty$ means $\lim_{t\to\infty}\frac{\vecx_t}{t}$ and $S^{d-1}$ is the $(d-1)$-surface of the $d$-dimensional unit sphere.
This mapping from $\vecx_0$ to the distribution of $\lim_{t\to\infty}\frac{\vecx_t}{t}$ serves as a {\it microscope,} which allows to assess $\vecx_0$  through the repeated measurements of $\vecx_t$ at sufficiently large $t.$
Notice that the mapping, $\vecx_0\mapsto Prob^{\rm (can)}(\cos\theta<\chi |\vecx_0),$ is asymptotically independent of the micro-macro ratios such as $\|\vecx_0\|/\| \vecx_t\|.$ This is in contrast to the case of assessing the initial position $\vecx_0$ of a Brownian particle from its position at a later time, $\vecx_t.$ In the latter case the signal-to-background ratio lessens with time. 
See Appendix~\ref{app:x0plusWt} for more explanation.

\section{\null{Discussion}} \label{sec:Discussion}
\null{The present work may bring two things.
First, we have found a connection between the Langevin equation and the Langevin functions through the martingale process. Usually the relationship between the Langevin equation and canonical distribution is through the Einstein's relation by which the stationary state of the Langevin equation become canonical (if the drift has a potential). In the present case, however, the evolution of neither $\vecx_t$ nor $\veca(\vecx_t)$ has {\it regular} stationary density; $\vecx_t$ grows unboundedly while $\veca(\vecx_t)$ is trapped asymptotically at a point on the surface of unit hyper-sphere.  Nevertheless, the canonical spin statistics, which could give the Langevin function as response, emerges in the distribution of asymptotic limit of $\vecx_t/t.$ In Appendix~\ref{app:caricature}
we present a qualitative parable to give a glimpse of our results.

Secondly, our report is another example in which the self-referential condition brings us a non-trivial outcome, like the celebrated diagonal arguments by Cantor and by G\"odel, the fractals, to mention but a few. Usually the martingale process is constructed from a reference process so that we can apply useful theorems of the martingale theory to the former, and eventually, we gain insights about the reference process. The present setup asks, instead, what type of reference process posesses the self-referential martingale property.

The underlying physics by which the martingale brought the Langevin function is unknown. That the linearising transformation for the Riccati equation, $L_d(x)=Z_d'(x)/Z_d(x),$ takes the form of the canonical response might be a clue to this question.  Also unknown is how the martingale constraint transmits the initial molecular field $\vecx_0$ to the asymptotic distribution of ``spin,'' 
$\veca^{\,*}_\infty.$
In short the present findings bring us questions, rather than answers, about the relevance of martingale in physics beyond being a mere mathematical tool. 
We, therefore, would like to conclude our report by several open questions:
Can the space-time harmonic function (see, for e.g., \S 3.2.5.3 of \cite{review300}) be also a fixed point? 
Can the microscope idea developed below (\ref{eq:propo}) be generalizable to non-martingale drift ?
Are there links to the stochastic thermodynamics \cite{LNP,Peliti-book,Shiraishi-book}?
Can there be a parallel framework for the state space having topologies other than Euclidean one?   Will there be a quantum counterpart? Does the self-harmonic drift optimize some physical entity or information ? }

\acknowledgements
The author acknowledges Charles Moslonka for the regular discussion on the martingale and the Progressive Quenching. The author gratefully appreciates the helpful comments on the early draft by Matteo Polettini and the insightful discussion with Felix Ritort.

\appendix

\section{Family of self-harmonic drifts obtained by rescaling} \label{app:scaling}
For different interpretations of $\vecx_t$ the self-harmonicity of the drift implies different constraints between the drift and noise terms.
For example, when $\vec{y}_t$ represents a position of a particle in a heat bath
of temperature $T$ and friction coefficient $\gamma,$ the choice $(\alpha,C)=(1,{2\mathcal{D}})$ with $\mathcal{D}=\frac{\kT}{\gamma}$  in (\ref{eq:scale}) leads to
\beq 
d\vec{y}_\tau = \frac{2\kT}{\gamma} L_d({\|\vec{y}_\tau\|}{}) \hat{y}_\tau d\tau + \sqrt{
\frac{2\kT}{\gamma}}\, d\vec{W}_\tau, 
\eeq
Since $ L_d({\|\vec{y}_\tau\|}{}) \hat{y}_\tau=\nabla \ln Z_d(\|\vec{y}\|),$
this equation describes the motion in the potential, $(-2\kT  \ln Z_d(\|\vec{y}\|)).$
To assure the self-harmonicity the energy scale of the potential is thus constrained.

Another case to apply (\ref{eq:scale}) is through an analogy with 
the Progressive Quenching of $d$-dimensional Heisenberg spins on a complete network \cite{PQ-KS-BV-pre2018,PQ-chain,PQ-CM-KS-2020,PQ-CM-KS-2022}, which we have mentioned in the last paragraph of Sec.\ref{sec:intro}. 
In this context the appropriate choice is $(\alpha,C)=(\frac{J}{\kT},\frac{\kT}{J}),$ which brings  (\ref{eq:scale})  into the following:
\beq\label{eq:eqLPQ}
d\vec{y}_\tau =  L_d\inRbracket{\frac{J\|\vec{y}_\tau\|}{\kT}}
 \hat{y}_\tau d\tau + \sqrt{\frac{\kT}{J}}\, d\vec{W}_\tau.
\eeq
In this picture  we regard $L_d\inRbracket{\null{J\|\vec{y}_\tau\|}/{\kT}}\hat{y}_\tau$ on the r.h.s. of (\ref{eq:eqLPQ}) as the equilibrium mean of a Heisenberg spin under the molecular field $J\vec{y}_\tau$ with $J$ being a spin-spin coupling constant and $\kT$ being the temperature.
Then a new fragment of spin $d\vec{y}_\tau$ appears at every interval $d\tau$  and joins the quenched part, $\vec{y}_\tau.$ While the conditional mean of $d\vec{y}_\tau$ follows the canonical statistics, there is  the Wiener noise $\propto d\vec{W}_\tau$ on top of it. To assure the self-harmonicity the amplitude of the noise must be of $\sqrt{\null{\kT}/{J}}.$

\section{Long-time convergence of $\vecx_t/t$} \label{app:convergence}
In the main text we see that, for every particular trajectory of $\veca^{\,*}(\vecx_t)$ 
starting from $x_t=x_0,$ there is a limit, $\veca^{\,*}_\infty$ with $\|\veca^{\,*}_\infty\|=1.$ 
Below we will see that the ratio, ${\vecx_t}/{t},$ also converges to $\veca^{\,*}_\infty.$ 

Once we admit the existence of the limit $\veca^{\,*}_\infty,$  
there exists a $u$ such that $\| \veca^{\,*}(x_t)-\veca^{\,*}_\infty\|<\epsilon$ for $\forall t \ge u$ for a given constant $\epsilon(>0).$
When we integrate (\ref{eq:eqSDE}) over the interval $[u,t]$ as 
\beq\label{eq:ballisticx}
\vecx_t=\vecx_u+\int_u^t \veca^{\,*}(\vecx_s)ds + \int_{s=u}^{s=t} dW_s,
\eeq
we see that $\vecx_u /t\sim t^{-1}$ and $\int_{s=u}^{s=t} dW_s /t\sim t^{-1/2}$ for $t\to \infty.$ We, therefore, focus on $ \int_u^t \veca^{\,*}(\vecx_s)ds,$ which reads
$$
\int_u^t \veca^{\,*}(\vecx_s)ds=(t-u)\veca^{\,*}_\infty+ \int_u^t [\veca^{\,*}(\vecx_s)-\veca^{\,*}_\infty]ds.
$$
While the magnitude of the integrand on the r.h.s.\,is already bounded by $\epsilon,$
we expect the integrand to decay almost like $s^{-1}$ because 
for large $\vecx_s$ the Langevin equation (\ref{eq:eqL}) means roughly
$d\vecx_s/ds\simeq (1-\mathcal{O}(\|\vecx_s\|)) \hat{x}_s+d\vec{W}_t,$
where we noticed $L_d(x) =1- \mathcal{O}(x^{-1})$ for $x\to\infty.$ 
Therefore, the time integral on the r.h.s. is dominated by $(t-u)\epsilon.$
Dividing each term in (\ref{eq:ballisticx}) by $t,$ we reach the claimed result;
$$
\lim_{t\to\infty}\frac{\vecx_t}{t} =\veca^{\,*}_\infty,
\mbox{ when } \lim_{t\to\infty}\veca^{\,*}(\vecx_t) =\veca^{\,*}_\infty.
$$

\section{Inferring the initial position of a Brownian particle a posteriori}\label{app:x0plusWt}
When a Brownian particle starts from $\vecx_0,$ its position at time $t,$ that is $\vecx_t=\vecx_0+\vec{W}_t,$ is martingale associated with the Wiener process $\vec{W}_t.$ While the mean $\la \vecx_t|\vecx_0\ra$ remains $\vecx_0,$ the variance $\la (\vecx_t-\vecx_0)^2|\vecx_0\ra$ grows linearly in time.
Therefore, unlike the case of self-harmonic drift, the inference of $\vecx_0$ through the observations of $\vecx_t$ at a ``macroscopic'' distance $R(\gg \|\vecx_0\|\ge 0)$ is hard to realize under the signal-to-background ratio decaying with $R.$ In fact, through the analysis of the exit problem \cite{FPT-KS}, the detected position on the circle, $\|\vecx\|=R,$ obeys the cumulative probability, 
{\small \beqa
Prob(\cos\theta<\chi)&=&
 \frac{1-\epsilon^2}{2}\inRbracket{\inv{\sqrt{1-2\epsilon \chi+\epsilon^2}}-\inv{1+\epsilon} } 
 \cr &=& \frac{1+\chi}{2}\inSbracket{1-\frac{3(1-\chi)}{2}\epsilon
 +\mathcal{O}(\epsilon^2)},
\cr &&
\eeqa}
where $\epsilon\equiv \|\vecx_0\| /R.$ We see that the signal, the term $-\frac{3(1-\chi)}{2}\epsilon,$ vanishes in the macroscopic limit, $\epsilon\to 0.$\\

\section{Parable of the contents of present work}\label{app:caricature}
Here we give a qualitative parable to give a glimpse of the system that we found and analyzed, in disregarding the logical order and some notations in the main text.

 Suppose there is a $d$-dimensional space with Euclidean metric in which the ``field'' is radially oriented outwards and the strength at the position $\vecx$ is simply $\vecx.$ 
A space-ship that travels in this space has a ``compass'' made of the 
 $d$-dimensional spin $\hat{S}$ with $\|\hat{S}\|=1$ that feels the field $\vecx$ in which the space-ship finds itself.
The thermal fluctuation of this spin is, however, so rapid that the navigator of the space-ship can only observe the thermal average, $\la \hat{S}\ra=L_d(\|\vecx \|)\hat{x},$ where $\hat{x}$ is the unit vector along $\vecx$ and $L_d(x)$ is the Langevin function that saturates to unity for $x\to\infty.$
Starting from $\vecx_0$ at time $t=0,$
the instantaneous speed of the space-ship at position $\vecx_t$ is the sum of the drift which is simply the mean spin, $\la \hat{S}_t\ra\, (=L_d(\|\vecx_t \|)\hat{x_t}),$ and the standard white-Gaussian noise, $\xi_t,$ i.e., the time derivative of a Wiener process $\vec{W}_t.$ 
After a large enough time, the space-ship is far from the origin and the mean speed $\vecx_t/t$ converges to a unit vector, which we denote by $\hat{S}_\infty.$ The first finding (the martingale property) is that the ensemble average of the asymptotic mean speed $\hat{S}_\infty$  over the trajectories from $\vecx_0$ is identical to the initial speed, that is, $\la \hat{S}_\infty\ra_{\vecx_0}=\la \hat{S}_{0}\ra_{\vecx_0},$ 
where the r.h.s. is the thermal average of the spin at $\vecx_0.$
The second finding is that the ensemble distribution of $\hat{S}_\infty$ is identical to the thermal distribution of $\hat{S}_{0}$ at the starting point $\vecx_0,$ the level of information which the navigator did could not access.

\bibliographystyle{apsrev4-2.bst}
\bibliography{ken_LNP_sar.bib}

\end{document}